\documentclass[conference,a4paper]{IEEEtran}
\IEEEoverridecommandlockouts
\usepackage{cite}
\usepackage{amsmath,amssymb,amsfonts}
\usepackage{algorithmic}
\usepackage{graphicx}
\usepackage{textcomp}
\usepackage{xcolor}
\def\BibTeX{{\rm B\kern-.05em{\sc i\kern-.025em b}\kern-.08em
    T\kern-.1667em\lower.7ex\hbox{E}\kern-.125emX}}
\begin{document}

\title{A Toolchain for Comprehensive Audio/Video Analysis Using Deep Learning \\ Based Multimodal Approach \\ (A use case of riot or violent context detection)
}

\author{\IEEEauthorblockN{Lam Pham*}
\IEEEauthorblockA{\textit{Austrian Institute of Technology} \\
Vienna, Austria}
\and
\IEEEauthorblockN{Phat Lam*}
\IEEEauthorblockA{\textit{HCM University of Technology} \\
HCM city, VietNam }
\and
\IEEEauthorblockN{Tin Nguyen*}
\IEEEauthorblockA{\textit{HCM University of Technology} \\
HCM city, VietNam} 
\and 
\IEEEauthorblockN{Hieu Tang}
\IEEEauthorblockA{\textit{FPT University} \\
HCM city, VietNam}
\and
\IEEEauthorblockN{Alexander Schindler}
\IEEEauthorblockA{\textit{Austrian Institute of Technology} \\
Vienna, Austria}

\thanks{(*) Lam Pham, Phat Lam, and Tin Nguyen made equal contribution to this paper.}%

}

\maketitle

\begin{abstract}
In this paper, we present a toolchain for a comprehensive audio/video analysis by leveraging deep learning based multimodal approach.
To this end, different specific tasks of Speech to Text (S2T), Acoustic Scene Classification (ASC), Acoustic Event Detection (AED), Visual Object Detection (VOD), Image Captioning (IC), and Video Captioning (VC) are conducted and integrated into the toolchain.
By combining individual tasks and analyzing both audio \& visual data extracted from input video, the toolchain offers various audio/video-based applications: Two general applications of audio/video clustering, comprehensive audio/video summary and a specific application of riot or violent context detection.
Furthermore, the toolchain presents a flexible and adaptable architecture that is effective to integrate new models for further audio/video-based applications.     
\end{abstract}


\section{INTRODUCTION}
\label{intro}
Recently, various tasks of video-based analysis have been proposed by the computer vision research community such as video captioning, activity detection, video similarity, human face recognition, etc. 
Generally, these tasks are specific and proposed together with certain benchmark datasets.
For instance, the tasks of video captioning, activity detection, video similarity, etc. have been evaluated with the large and benchmark datasets of TVR~\cite{vc_ds}, ActitityNet~\cite{act_ds}, FIVR~\cite{fivr_ds}, respectively.
Additionally, models proposed for these video-based analysis tasks have leveraged a wide range of deep neural networks and deep learning techniques.
For example, SwinBBERT, a variant of transformer architecture, was proposed for the task of video captioning~\cite{vc_model}.
Meanwhile, a CNN-based architecture was proposed for the task of video similarity~\cite{video_emb}.
Therefore, these models inevitably are overfitting in certain datasets and only suitable for specific tasks defined in advance.
This leads to be impossible to achieve a single model which can adapt to different video-based analysis tasks.
%
\begin{figure*}[t]
    \centering
    \includegraphics[width = 1.0\linewidth]{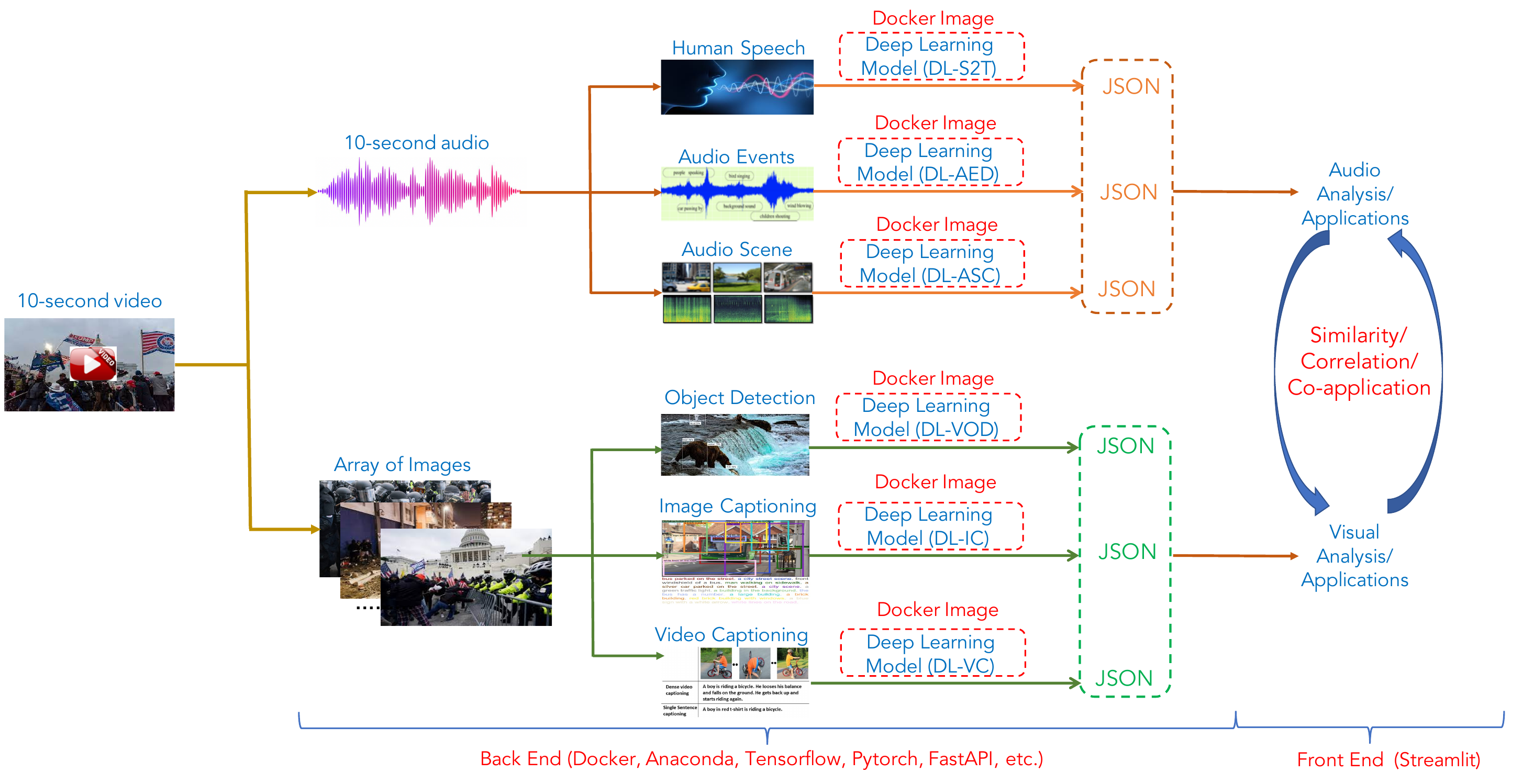}
    	\vspace{-0.2cm}
	\caption{The high-level architecture of the proposed toolchain for comprehensive audio/video analysis}
    \label{fig:f1}
\end{figure*}

Regarding two main information of visual data (e.g. array of images) and audio data extracted from a video, video-based applications mainly focus on exploring visual information~\cite{video_emb}.
This can be explained that the visual data extracted from the video contains rich information rather than audio data for a wide range of video-based applications such as activity detection, human face recognition, traffic monitoring, etc.
However, analyzing audio data is important and essential for certain applications such as gun-relevant event detection or violent-relevant context detection~\cite{lampham_01}.
Even with the task of television programme classification which conventionally explores visual information, only exploring audio helps to achieve competitive results in~\cite{bbc_pp}.

Given the analysis above, it can be seen that achieving a general and comprehensive audio/video analysis requires: (1) Multimodal approach which combines multiple individual models and (2) Analyzing both visual and audio data extracted from input videos.
Indeed, recently proposed video-based systems have analyzed both audio and visual information and leveraged individual deep learning based models.
For example, the task of emotion detection proposed in~\cite{mm_emotion} presents an effective combination of individual tasks: text-based emotion detection (e.g. the content of human speech), audio-based emotion detection (e.g. the tone of human speech) , and visual-based emotion detection (e.g. the human face) . 
Similarly, authors in~\cite{lampham_03} proposed a system for daily life video classification.
This work proved that exploring both audio and visual data is effective to enhance the task performance.

Motivated by this research trend, we develop and present a toolchain for a comprehensive audio/video analysis which leverages deep learning based and multimodal approaches in this paper.  
The toolchain, which integrates a wide range of individual tasks of Speech to Text (S2T), Acoustic Scene Classification (ASC), Acoustic Event Detection (AED), Visual Object Detection (VOD), Image Captioning (IC), and Video Captioning (VC), offers two general applications of audio/video clustering and comprehensive audio/video summary and a specific audio/video-based application of riot or violent context detection.
 \begin{table}[t]
	\caption{Individual models integrated in the toolchain} 
	\vspace{-0.1cm}
	\centering
    \scalebox{0.8}{
	\begin{tabular}{|c |c|c| } 
		\hline 
		\textbf{Models}   &\textbf{Tasks} &\textbf{Licenses}  \\ 
		\hline 
    	\hline 
		Whisper~\cite{whisper}, &Speech to Text (S2T), & MIT   \\        
		                  mBart~\cite{mbart_model} &English Translation (ET) &MIT            \\
		    	\hline 
		PANN~\cite{kong_pretrain} &Acoustic Event Detection (AED)  &MIT     \\    
		    	\hline 
     	AIT-ASC~\cite{lampham_01, lampham_02} &Acoustic Scene Classification (ASC) & \textbf{Our development}     \\        
     	    	\hline 
     	DETR~\cite{vod_model} &Visual Object Detection (VOD) &Apache-2.0        \\        
     	     	    	\hline 
     	VEDA~\cite{ic_model} &Image Captioning (IC) &Apache-2.0        \\        
     	     	    	\hline 
     	SWINBERT~\cite{vc_model} &Video Captioning (VC) &MIT       \\             	     	     	     	
		\hline 
        \end{tabular}  
	}  
	\label{table:res1} 
\end{table}
\section{The Toolchain architecture}
\label{frameworks}
Figure~\ref{fig:f1} presents the high-level architecture of our proposed toolchain for the comprehensive audio/video analysis.
First, audio data and visual data (images) are extracted from an input video.
Regarding audio data, it contains three main audio information: human speech, audio events (e.g. gun sound, dog bagging, etc. occurring in the audio recording), and audio scene (e.g. background noise in a pedestrian street, in a music event, in sport atmosphere, etc.).
For each type of audio information, a deep learning based model is proposed to explore and then export the result into a JSON file.
These JSON files store the text of human speech (the task of Speech to Text (S2T)), the probability and name of audio events occurring (the task of Audio Event Detection (AED)), and the probability and name of audio scene (the task of Audio Scene Classification (ASC)).

Similarly, three main tasks for analyzing visual data (images) are defined: Visual object detection (VOD), Image Captioning (IC), and Video Captioning (VC). 
Each task is solved by individual deep learning based models and also reports the results in a JSON file.
These JSON files store the names of visual objects, the description of images, and the description of the video, respectively.
Finally, the JSON files, which present various audio and visual information in text, are analyzed and used for three tasks in this paper: (1) audo/video clustering; (2) comprehensive audio/video summary; and (3) riot or violent context detection.

Deep learning models and their corresponding licenses used to solve individual tasks are presented in Table~\ref{table:res1}.
We use FastAPI framework for the backend development while Streamlit is for the frontend development.
Notably, each specific deep learning model, which was constructed in individual Anaconda environment with Tensorflow or Pytorch framework, is packaged by Docker container. 
By constructing and packaging single models for specific tasks independently, the toolchain architecture shows effectiveness and adaptability to integrate new tasks.
\subsection{Datasets for simulation}
\label{dataset}
To evaluate and simulate the proposed toolchain, we use three benchmark datasets of VCDB~\cite{vcdb_ds}, DCASE 2021 Challenge Task 1B~\cite{ds_2021_1b}, and Crowded Scene~\cite{lampham_04}.
While DCASE 2021 Task 1B dataset presents 10 video scenes of daily life (airport (indoor), shopping mall (indoor), metro station (indoor), pedestrian street (outdoor), public square (outdoor), pedestrian street (outdoor), park (outdoor), tram (transportation), bus (transportation) and metro (transportation)), the Crowded Scene dataset presents 5 very noise scenes (firework, riot, very noise street, music event, sport atmosphere).
Meanwhile, VCDB data presents different video events collected from various resources of films and social media (e.g. YouTube, etc.).
\begin{figure}[t]
    \centering
    \includegraphics[width = 0.9\linewidth]{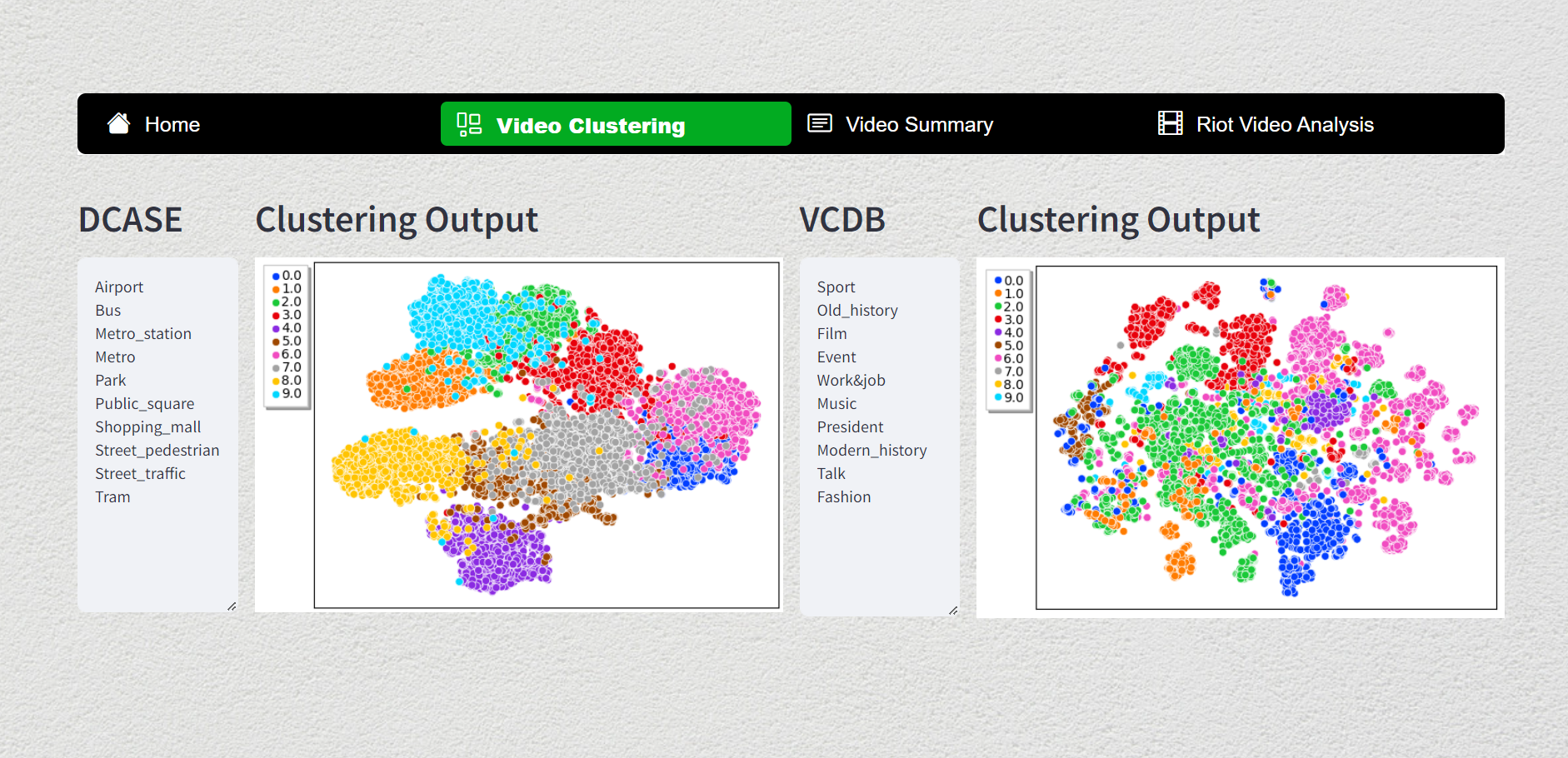}
    	\vspace{-0.2cm}
	\caption{The simulation of audio/video clustering}
    \label{fig:f2}
\end{figure}

\section{Simulation and Visualization}
\label{result}

\textbf{Audio/video clustering:} Given the amount of videos from various resources (i.e. Around 3.7 million new videos are uploaded to YouTube everyday), clustering videos into certain categories is necessary before further analyzing the video content. Therefore, the proposed toolchain first  offers the audio/video clustering application.
To this end, audio and visual data extracted from input videos are fed into pre-trained models to extract audio and visual embeddings. 
These embeddings are transformed into two-dimensional vectors using TSNE and then visualized.
This application is evaluated by using datasets of VCDB~\cite{vcdb_ds}, DCASE 2021 Challenge Task 1B~\cite{ds_2021_1b}.
As Figure~\ref{fig:f2} shows, videos from DCASE 2021 Task 1B and VCDB are classified into 10 different topics corresponding to 10 different colors. 

\begin{figure}[t]
    \centering
    \includegraphics[width = 0.9\linewidth]{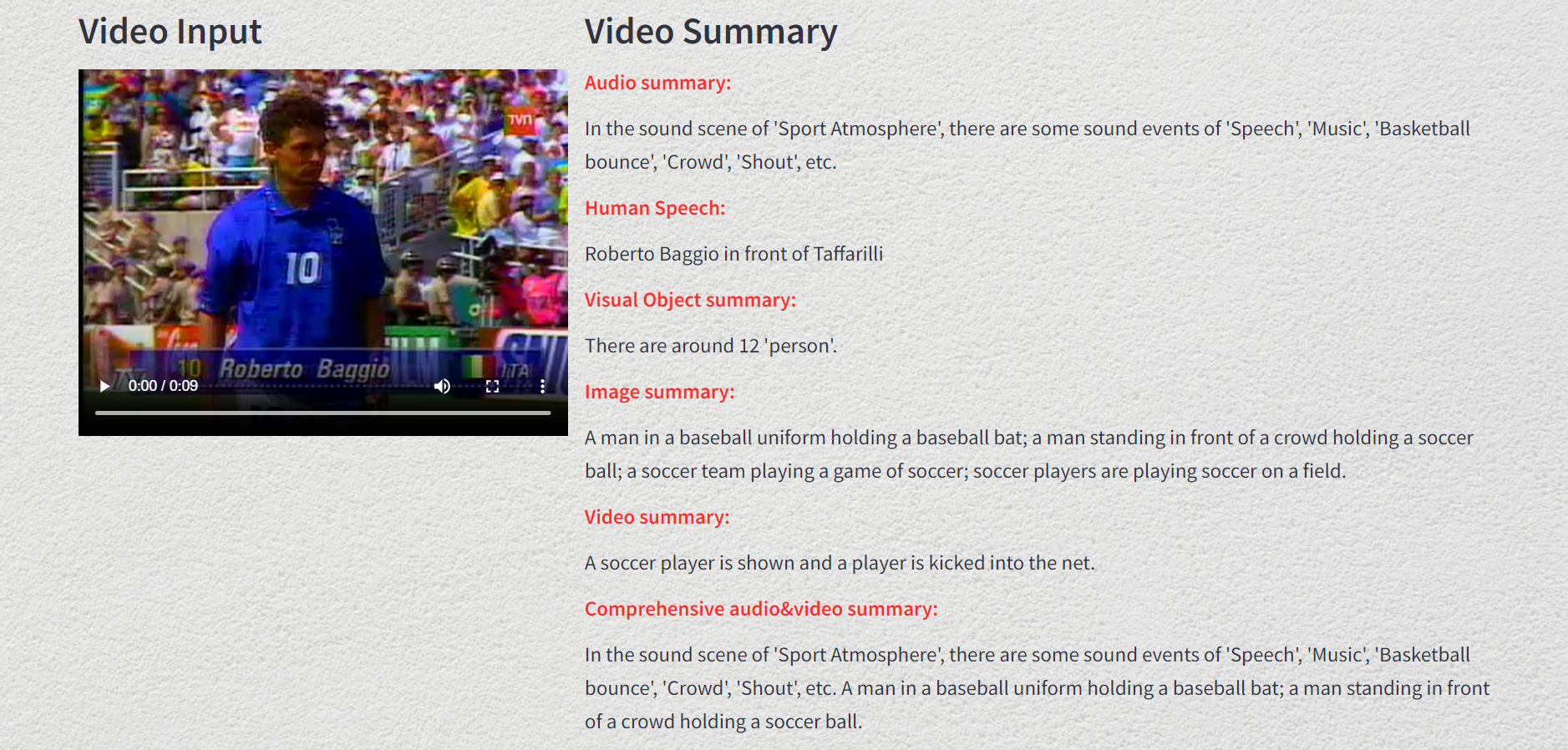}
    	\vspace{-0.2cm}
	\caption{The simulation of audio/video summary}
    \label{fig:f3}
\end{figure}

\textbf{Comprehensive audio/video summary:}
To further analyze a certain video from a video group, we analyze text data obtained from individual audio and visual based models.
As Figure~\ref{fig:f3} shows, the football video (i.e. a video sample from VCDB dataset) is presented in text description regarding different audio and visual aspects.
The audio events of `Speech', `Music', `Basketball bounce', `Crowd', `Shout' and the audio scene of `Sport Atmosphere' explain that a very noisy and sport-relevant context is happening in the input video.
Meanwhile, a lot of visual objects of `person', the image captioning of `a crowd of people watching a soccer game', and the video captioning of `a soccer game is being played and a crowd cheers' further support the audio-based context and strongly confirm a sport event. 

\begin{figure*}[t]
    \centering
    \includegraphics[width = 0.95\linewidth]{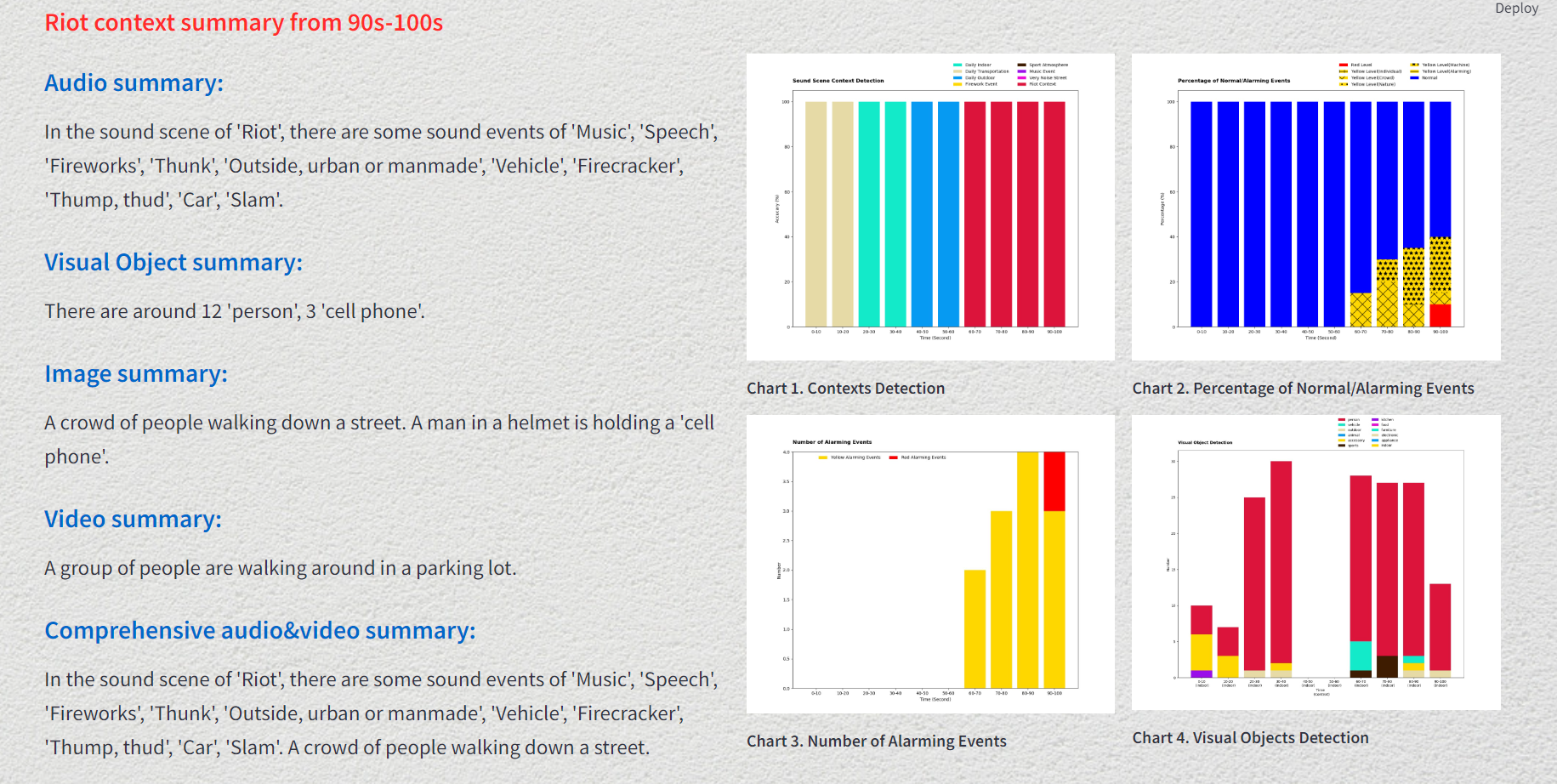}
    	\vspace{-0.2cm}
	\caption{The simulation of riot or violent context detection}
    \label{fig:f4}
\end{figure*}

\textbf{Riot or violent context detection:}
Given two applications above, we now can classify and summarize a certain input audio/video.
These applications help to provide general and necessary information for further analyzing an input video.
By leveraging the information, a specific application of riot or violent context detection is proposed.
To this end, keywords of `gun', `scream', `crowd', etc., which are  related to the riot or violent context, are defined in advance.
These keywords are then compared with the given information from the task of audio/video summary to indicate whether the video is close to a riot or violent context.
The keywords are separated into three alarming levels which correspond to blue, yellow, and red colors, respectively.
The keywords such as `speech', `car', etc. which are marked by the blue level present events occurring in normal daily life.
Meanwhile, the yellow keywords such as `crowd', `scream', etc. and  the red keywords of `gun', `explosion', etc. present a middle and high alarming situation, respectively.
To simulate the riot or violent context detection, we create an input video which presents the change of scene contexts: Metro (0 second to 20 second), Metro-Station (20 second to 40 second), Park (40 second to 60 second), and finally Riot context (60 second to 100 second).
Figure~\ref{fig:f4} presents our simulation results of the riot/violent context detection. 
The text summary in Figure~\ref{fig:f4} presents the riot context occurring with the sound events, the sound scenes, the visual objects, and the context description.
The change of scene contexts is presented in the chart 1 which is the result of the acoustic scene classification task (ASC).
The riot/violent-relevant sound events corresponding to three alarming levels are presented in the chart 2.
When the riot occurs from 60 second, distinct sound events of `Firework', `Firecracker', `Thump, thud', etc. are detected. 
The red color helps to indicate and raise a dangerous situation. 
The density of yellow and red sound events in the chart 3 presents the severity of the situation.
The chart 4 indicates that the context is related to human activity (i.e. `person' objects are marked with the red color) with a high density of human-relevant keywords detected.
By combining text-based audio/video summary and doing statistics on defined keywords, a riot or violent context can be detected, observed, and estimated the severity effectively.

\section{Conclusion}
\label{conclusion}
We have presented a toolchain for comprehensive audio/video analysis using deep learning based multimodal approach.
As the toolchain integrates a wide range of individual models for specific tasks and explores audio \& visual data, the toolchain offers two general applications of audio/video clustering and comprehensive audio/video summary.
Given these general applications, a specific application of riot or violent context detection is then developed with a definition of keywords related to the context.
Flexibly conduct a specific task (e.g. riot or violent context detection) based on the two general tasks (e.g. audio/video clustering, comprehensive audio/video summary) and the adaptable toolchain architecture allow the integration of a wide range of new applications such as crowd detection, music event detection, anomaly detection, specific human activity detection, etc. only with relevant keywords defined in advance.
For future work, a text-based and query-answer method will be developed which prompts a text query to find videos presenting the most relevant context.
This method also leverages two general applications of audio/video clustering and comprehensive audio/video summary proposed in this paper.



\bibliographystyle{IEEEtran}
\bibliography{refs}

\end{document}